\documentclass[utf8,sort&compress]{frontiersFPHY}

\usepackage{color}
\usepackage{amsmath,amsfonts,amsthm,amssymb}
\usepackage{bm,bbm}
\usepackage[OT2,OT1]{fontenc}

\usepackage{url,hyperref,lineno,microtype,subcaption}
\usepackage[onehalfspacing]{setspace}

\newcommand{\be}{\begin{equation}}
\newcommand{\ee}{\end{equation}}
\newcommand{\ba}{\begin{eqnarray}}
\newcommand{\ea}{\end{eqnarray}}
\def\bs{\begin{subequations}}
\def\es{\end{subequations}}
\renewcommand{\leq}{\leqslant}
\renewcommand{\geq}{\geqslant}
\def\a{\alpha}
\def\b{\beta}
\def\de{\delta}

\def\om{\omega}

\def\vr{\varrho}

\def\cD{\mathcal{D}}

\def\cK{\mathcal{K}}

\def\p{\partial}
\def\bp{\bar{\partial}}
\def\B{\Box}
\newcommand{\Eq}[1]{(\ref{#1})}

\def\cob{\color{blue}}
\newcommand{\book}[5]{\emph{#1}; #2: #3, #4, #5}
\newcommand{\books}[4]{\emph{#1}; #2: #3, #4}
\newcommand{\oarX}[1]{\href{http://arxiv.org/abs/#1}{{\ttfamily\cob arXiv:#1}}}
\newcommand{\arX}[1]{\href{http://arxiv.org/abs/#1}{{\ttfamily\cob arXiv:#1}}}
\newcommand{\doin}[6]{\href{http://dx.doi.org/#1}{{\cob {\it #2 #3} (#6) {\bf #4}:#5}}}
\newcommand{\doinn}[5]{\href{http://dx.doi.org/#1}{{\cob {\it #2} (#5) {\bf #3}:#4}}}
\newcommand{\doij}[5]{\href{http://dx.doi.org/#1}{{\cob {\it #2} (#5) {\bf #3}:#4}}}

\newcommand{\ndoinn}[5]{\href{#1}{{\cob {\it #2} (#5) {\bf #3}:#4}}}
\newcommand{\tia}[1]{#1.}

\def\rmd{d}

\def\bd{\mathbbm{d}}
\newcounter{listcounter}

\def\keyFont{\fontsize{8}{11}\helveticabold }
\def\firstAuthorLast{Gianluca Calcagni}
\def\Authors{Gianluca Calcagni\,$^{*}$}
\def\Address{Instituto de Estructura de la Materia, CSIC, Serrano 121, 28006 Madrid, Spain}
\def\corrAuthor{Gianluca Calcagni}

\def\corrEmail{g.calcagni@csic.es}

\begin{document}
\onecolumn
\firstpage{1}

\title[Towards multifractional calculus]{Towards multifractional calculus} 

\author[\firstAuthorLast]{\Authors} 
\address{} 
\correspondance{} 

\extraAuth{}

\maketitle

\begin{abstract}
After motivating the need of a multiscale version of fractional calculus in quantum gravity, we review current proposals and the program to be carried out in order to reach a viable definition of scale-dependent fractional operators. We present different types of multifractional Laplacians and comment on their known or expected properties.

\tiny
 \keyFont{\section{Keywords:} quantum gravity, fractional calculus, fractional derivatives, multiscale geometry, multifractional spacetimes} 
\end{abstract}

\date{\small December 31, 2017}


\section{Introduction}

A branch of theoretical physics which has been attracting considerable attention in the last years is quantum gravity. Several independent theories, models and hypotheses are gathered under this broad name, from string theory to asymptotic safety, from nonlocal to loop quantum gravity, from causal dynamical triangulations to causal sets, and so on \cite{Ori09,Fousp,CQC,MiTr}. Most of these proposals aim to conciliate classical general relativity with the laws of quantum mechanics, in order to unify all forces of Nature under the same framework and to solve some problems left open in the traditional paradigms (for instance, the big-bang and cosmological constant problems \cite{CQC}).

A surprising feature emerging from this variegated landscape is that the properties of spacetime geometry, such as the spectral or Hausdorff dimension and the way particles diffuse, change with the probed scale in all quantum gravities \cite{tH93,Car09,Car17}. This so-called dimensional flow seems to be a manifestation of the impossibility to perform infinitely precise time and distance measurements in geometries with intrinsic uncertainties of quantum or stochastic origin \cite{ACCR,CaRo2}. Some of these findings were made possible by assuming dimensional flow by default and treating spacetime geometry as fundamentally scale dependent. This general method can be embodied in a class of theories, called \emph{multifractional}, where classical and quantum fields live on a spacetime characterized by a scale hierarchy, anomalous transport and correlation properties, and a multifractal geometry \cite{revmu}. Surprisingly, all these features emerge automatically by assuming a slow dimensional flow at large scales (dimension in the infrared almost constant) \cite{revmu,first}.

One can encode a multiscale geometry in the dynamics of particles and fields in several ways. The one followed by multifractional theories is a change in the integro-differential structure \cite{frc1}. Integrals (such as dynamical actions) and derivatives (in kinetic terms) acquire a nontrivial scale dependence that can be illustrated in the prototype example of the scalar field theory
\be
S=\int\rmd\vr(x)\,\left[\frac12\phi\cK\phi-V(\phi)\right],
\ee
where $\vr(x)$ is the spacetime measure, $\cK$ is a kinetic operator, and $V$ is the scalar potential. In the standard case and in the absence of gravity (which will be ignored here), $\vr(x)=\rmd^Dx$ is the usual Lebesgue measure in $D$ topological dimensions and $\cK=\B=\p_\mu\p^\mu$ is the second-order Laplace--Beltrami operator. In the presence of dimensional flow, if the measure is factorizable in the coordinates (an assumption to make the problem tractable) then it takes the unique form \cite{first}
\be\label{qom}
\vr(x)=\prod_\mu \rmd q^\mu(x^\mu)\,,\qquad q^\mu(x^\mu)= x^\mu+\sum_{n=1}^{+\infty} \frac{\ell_n^\mu}{\a_{\mu,n}}{\rm sgn}(x^\mu)\left|\frac{x^\mu}{\ell_n^\mu}\right|^{\a_{\mu,n}} F_n(x^\mu)\,,
\ee
where 
\be\label{Fom}
F_\om(x^\mu)= 1+A_{\mu,n}\cos\left(\om_{\mu,n}\ln\left|\frac{x^\mu}{\ell_{\infty,n}^\mu}\right|\right)+B_{\mu,n}\sin\left(\om_{\mu,n}\ln\left|\frac{x^\mu}{\ell_{\infty,n}^\mu}\right|\right)\,,
\ee
all indices $\mu$ are inert (there is no Einstein summation convention), the first factor 1 in \Eq{Fom} is optional \cite{first} (it can be set to zero in the stochastic version of the theory \cite{CaRo2}), $\ell_n^\mu$ and $\ell_{\infty,n}^\mu$ are $2D$ length scales for each $n$, and $\a_{\mu,n}$, $A_{\mu,n}$, $B_{\mu,n}$, and $\om_{\mu,n}$ are $4D$ real constants for each $n$.

Since the measure is factorized, in the following we can focus the discussion on the one-dimensional model
\be\label{S1d}
S=\int\rmd q(x)\,\left[\frac12\phi\cK\phi-V(\phi)\right],\qquad q(x)\simeq x+\frac{\ell_*}{\a}{\rm sgn}(x)\left|\frac{x}{\ell_*}\right|^{\a} F_\om(x)\,,
\ee
where $\ell_*=\ell_1$, all $\mu$ indices are omitted and we can also ignore terms subleading both in the infrared and in the ultraviolet; this corresponds to consider only the $n=1$ term in Eqs.\ \Eq{qom} and \Eq{Fom}. 

In this paper, we will study the properties of three versions of the kinetic operator $\cK$, expanding on the proposals sketched in \cite{revmu}. Since, in the context of quantum gravity, the integration measure is uniquely defined independently of the type of derivatives in the Lagrangian as in Eq.\ \Eq{S1d} \cite{first}, here we are not interested in the formal properties of ``multiscale integrals,'' the inverse of multiscale derivatives. Some of these operators are known, as is the case of \Eq{qder} and \Eq{capuq2} below \cite{revmu}, while in the case of \Eq{lio2} and \Eq{wey2} they are unknown and will require further work. On the other hand, there is no inverse operator for a linear combination of operators with different inverse, such as Eq.\ \Eq{exms}. In all these cases, for our purposes it is sufficient to study the properties of multiscale derivatives with respect to the ordinary Lebesgue measure $\rmd x$ while, at the same time, taking into account the measure weight by inserting weight factors in the definitions of such derivatives to make them self-adjoint with respect to the measure.


\section{\texorpdfstring{$q$}{}-derivatives}

While there is a unique parametric form of the measure $q(x)$, there is more freedom in the choice of kinetic operator $\cK$. It turns out that there are three viable possibilities. One is a theory with so-called weighted derivatives, but this can be reduced to a system with ordinary derivatives and the spectral dimension of spacetime is constant in that case \cite{revmu}. Another possibility is the second-order operator \cite{frc2}
\be\label{qder}
\cK=\p_q^2\,,\qquad \p_q:=\frac{\p}{\p q(x)}=\frac{1}{v(x)}\frac{\p}{\p x}\,,
\ee
where $v(x)=q'(x):=\p_x q(x)$ and $q(x)$ is given by Eq.\ \Eq{qom}. This ``$q$-derivative'' has a number of highly desirable properties:
\begin{enumerate}[leftmargin=*,labelsep=4.9mm]
\item	It is multiscale, since the scale hierarchy is already encoded in the measure weight $v(x)$.
\item	Its composition law is very simple:
\be
\p_q^2:=\p_q\p_q=(\p_q)^2-\frac{v'}{v^3}\p_x\,.
\ee
\item It is linear. For any $f$ and $g$ in a suitably defined functional space,
\be
\p_q(f+g)=\p_q f+\p_q g
\ee
\item Its kernel is trivial and given by a constant:
\be\label{ker}
\p_q 1=0\,.
\ee
\item	The Leibniz rule is extremely simple. For any $f$ and $g$ in a suitably defined functional space,
\ba
\p_q (fg) &=& \frac{1}{v}(f'g+fg')\nonumber\\
&=& (\p_q f)\,g+f\,(\p_q g)\,.\label{Leib}
\ea
\item Integration by parts is straightforward. For any $f$ and $g$ in a suitably defined functional space,
\ba
\int\rmd q\,f\,\p_q g &\stackrel{\textrm{\small\Eq{Leib}}}{=}& \int^{+\infty}_{-\infty}\rmd x\,v\,\frac{1}{v}(fg)'-\int\rmd q\,(\p_q f)\,g\nonumber\\
&=&-\int\rmd q\,(\p_q f)\,g\,,
\ea
where we threw away boundary terms. Consequently, $\cK$ is self-adjoint:
\be
\int\rmd q\,f\,\p_q^2 g=\int\rmd q\,(\p_q^2f)\,g\,.
\ee
\end{enumerate}
Notice that, in principle, these rules hold for an arbitrary $q(x)$, although in our case this profile is fixed as in \Eq{S1d}.


\section{Fractional derivatives}

The third extant multifractional theory is the least explored, but also the most interesting because it employs fractional calculus. This is by far the most obvious tool to implement an anomalous scaling in the geometry. The application of fractional derivatives to multiscale theories is not an easy task. Before seeing why, let us recall some basic aspects of fractional calculus. 

There are different versions of fractional derivatives \cite{MR,Pod99,KST}\footnote{For bibliographic references, see \cite{frc1,revmu}. More recent applications and solving methods can be found in \cite{BJAB,YTMB,SHZB,BIYA1,BIYA2,IYAB}.} and one must make a choice suitable for quantum gravity \cite{frc1}. In particular, we believe that one cannot renounce to have a trivial kernel (Eq.\ \Eq{ker}). Two fractional derivatives with this property are the Liouville derivative
\be\label{lio}
{}_\infty\p^\a f(x) := \frac{1}{\Gamma(m-\a)}\int_{-\infty}^{+\infty}\rmd x'\, \frac{\theta(x-x')}{(x-x')^{\a+1-m}}\p_{x'}^m f(x')\,,\qquad m-1\leq \a<m\,,
\ee
and the Weyl derivative
\be\label{wey}
{}_\infty\bar\p^\a f(x) := \frac{1}{\Gamma(m-\a)}\int_{-\infty}^{+\infty}\rmd x'\, \frac{\theta(x'-x)}{(x'-x)^{\a+1-m}}\p_{x'}^m f(x')\,,\qquad m-1\leq \a<m\,,
\ee
where $\theta$ is Heaviside's step function. Obviously, these operators act linearly on $f$ and ${}_\infty\p^\a_x 1= 0={}_\infty\bp^\a 1$. One can also check that ${}_{\infty}\p^\a{}_{\infty}\p^\b={}_{\infty}\p^{\a+\b}$ and ${}_{\infty}\bp^\a{}_{\infty}\bp^\b={}_{\infty}\bp^{\a+\b}$ (these fractional derivatives commute) and that the Leibniz rule is
\be\label{leru}
{}_\infty\p^\a(fg)=\sum_{j=0}^{+\infty}\binom{\a}{j} (\p^j f)({}_\infty\p^{\a-j}g)\,,\qquad \binom{\a}{j}=\frac{\Gamma(1+\a)}{\Gamma(\a-j+1)\Gamma(j+1)}\,,
\ee
and the same expression for the Weyl derivative, where $\p^{\a-j}=I^{j-\a}$ are integrations for $j\geq 1$. Also, integration by parts with the Liouville derivative generates the Weyl derivative, and vice versa: 
\be\label{ibp}
\int_{-\infty}^{+\infty}\rmd x\, f\,{}_{\infty}\p^\a g = \int_{-\infty}^{+\infty}\rmd x\, ({}_{\infty}\bp^\a f)\,g\,.
\ee


\subsection{Complicated Leibniz rule}

The importance to have the standard Leibniz rule \Eq{Leib} can be appreciated when trying to do physics with fractional calculus. In the theory with $q$-derivatives, integration by parts does not produce extra contributions and the kinetic terms $\int\rmd q\,\phi\p_q^2\phi$ or $-\int\rmd q\,\p_q\phi\p_q\phi$ are completely equivalent. Therefore, the equation of motion $\p_q^2\phi-V_{,\phi}=0$ can be determined easily by applying the variational principle on \Eq{S1d}. On the other hand, suppose we choose another type of multiscale derivative $\cD$ such that $\cK=\cD^2$ and its Leibniz rule is more complicated:
\be\label{Leib2}
\cD(fg) = (\cD f)\,g+f\,(\cD g)+X\,,
\ee
where $X=X(f,g;x)$ is a function of $f$, $g$, their ordinary derivatives and the coordinate $x$. For consistency, if the kernel of $\cD$ is trivial ($\cD 1=0$), then $X(f,1;x)=X(1,f;x)=0$ for any $f$. In particular, if $g=\cD h$, then
\ba
f\cD^2 h&=&\cD(f\cD h)-X(f,\cD h;x)-(\cD f)\,\cD h\nonumber\\
&=&[\cD(f\cD h)-\cD(h\cD f)-X(f,\cD h;x)+X(\cD f,\cD h;x)]+(\cD^2 f)\,h\nonumber\\
&=:& Y(f,h;x)+(\cD^2 f)\,h\,.\label{use}
\ea
Therefore, when varying the action \Eq{S1d} with respect to $\phi$ one gets
\ba
\frac{\de S}{\de\phi}&=&\int\rmd q\left(\frac12\de\phi\cD^2\phi+\frac12\phi\cD^2\de\phi-\de\phi V_{,\phi}\right)\nonumber\\
&\stackrel{\text{\small\Eq{use}}}{=}& \int\rmd q\,\left[\de\phi(\cD^2\phi-V_{,\phi})+\frac12Y(\phi,\de\phi;x)\right]\,.
\ea
Assuming that one could repeatedly integrate $Y$ by parts to write it as $Y=2\de\phi\,Z(\phi,x)$ up to some boundary term, we would end up with a dynamical equation
\be
\cD^2\phi-V_{,\phi}+Z(\phi,x)=0
\ee
characterized by a term $Z$ that can considerably hinder the study of solutions.

This is the main obstacle that prevented so far to consider multiscale theories with derivatives different from Eq.\ \Eq{qder} (barring the mathematically trivializable case of weighted derivatives). In fact, the only derivative with anomalous scaling such that $X=0$ in the Leibniz rule \Eq{Leib2} is the $q$-derivative \cite{Tar13}. Genuine fractional derivatives always have $X\neq 0$.


\subsection{Self-adjoint Laplacian}

Although $X\neq 0$, one could still obtain a clean integration by parts if, thanks to miraculous cancellations, $Y$ were a total derivative or $Z$ were zero on shell. This possibility is suggested by Eq.~\Eq{ibp}, which implies that, for any combination
\be\label{tilD0}
\tilde\cD^\a:=c\,{}_\infty\p^\a+\bar c\,{}_\infty\bp^\a,
\ee
one has
\be
\int\rmd x\, f\,(c\,{}_\infty\p^\a+\bar c\,{}_\infty\bp^\a) g =\int\rmd x\, g(\bar c\,{}_\infty\p^\a+c\,{}_\infty\bp^\a) f\,.\nonumber
\ee
For instance, if $c=-\bar c=1/2$,\footnote{A sign error in a similar expression in \cite{revmu} is corrected here.}
\be
\int\rmd x\, f\,\tilde\cD^\a g =-\int\rmd x\, (\tilde\cD^\a f)\,g\,,\qquad \tilde\cD^\a=\frac12({}_\infty\p^\a-{}_\infty\bp^\a)\,.\label{tilD}
\ee
In the limit $\a\to 1$, ${}_\infty\p^1=\p$ and ${}_\infty\bp^1=-\p$, so that $\lim_{\a\to 1}\tilde\cD^\a=\p$. Therefore, we can define an operator self-adjoint with respect to any measure weight $v(x)$:
\be
\cK_\a = \cD^\a\cD^\a\,,\qquad \cD^\a:=\frac{1}{\sqrt{v}}\tilde\cD^\a\left(\sqrt{v}\,\,\cdot\,\right)\,,\label{Ka}
\ee
so that
\ba
\int\rmd x\,v\,f\cK_\a g &=& \int\rmd x\, (\sqrt{v}f)\tilde\cD^\a\tilde\cD^\a(\sqrt{v}g)\nonumber\\
												 &=& -\int\rmd x\,[\tilde\cD^\a(\sqrt{v}f)]\tilde\cD^\a(\sqrt{v}g)\nonumber\\
												 &=& \int\rmd x\,[\tilde\cD^\a\tilde\cD^\a(\sqrt{v}f)](\sqrt{v}g)\nonumber\\
												 &=& \int\rmd x\,v\,(\cK_\a f)\,g\,.
\ea
Note that other, complex-valued choices of $c=(\bar c)^*$ may be more convenient when studying the spectrum of eigenvalues of these operators \cite{frc4}.


\section{Multifractional derivatives: three proposals}

At this point, we can try to extend fractional calculus to a multiscale setting. We have found three ways to do that.


\subsection{Explicit multiscaling}

The most direct mean to induce a hierarchy of scales and a variable anomalous scaling is to consider a superposition of fractional derivatives of different order $\a$ \cite{frc4}. In the mathematical literature, several authors \cite{Cap69,Cap95,BT1,BT2,CGS,LoH02,Koc1,Koc2} did propose a continuous superposition, the distributed-order fractional derivatives ${\rm D}:=\int_0^1\rmd\a\,m(\a)\,\p^\a$, where $m(\a)$ is a distribution on the interval $[0,1]$. However, from previous experience in quantum gravity it may be more convenient, or just sufficient, to take a sum instead of an integral:
\be\label{exms}
\cK=\cD^2\,,\qquad \cD:=\sum_n g_n\cD^{\a_n}\,,
\ee
where $g_n=g_n(\ell_n)$ are some constant coefficients and $\cD^{\a_n}$ is defined in Eqs.\ \Eq{Ka} and \Eq{tilD}. A nontrivial dimensional flow is generated by just one scale, i.e., a sum of two terms: $\cD=\p+g_*\cD^\a$. The equation of motion from the action \Eq{S1d} with kinetic operator \Eq{exms} is
\be\label{cd2}
\cD^2\phi-V_{,\phi}=0\,.
\ee

This formulation of a multiscale theory with fractional derivatives is not exempt from problems. The operator $\cD^2$ consists of many terms, even in the simplest case of only one scale where $\cD^2$ is made of seven pieces (ignoring weight factors), $\p^2+2g_*\tilde\cD^{\a+1}+g_*^2\tilde\cD^\a\tilde\cD^\a=\p^2+g_*({}_\infty\p^{\a+1}-{}_\infty\bp^\a)+(g_*^2/4)({}_\infty\p^{2\a}-{}_\infty\p^\a{}_\infty\bp^\a-{}_\infty\bp^\a{}_\infty\p^\a+{}_\infty\bp^{2\a})$. Therefore, the dynamics \Eq{cd2} is deceptively clean and hides a rather messy multiorder fractional differential structure which may be very difficult to solve analytically. This eminently practical issue could be very important, or even fatal, at the time of studying the dynamics. To bypass it, one could consider another version of the kinetic operator \cite{frc4}:
\be\label{exms2}
\cK=\sum g_n\bar\cD^{2\a_n}\,,\qquad \bar\cD^{2\a_n}:=\frac12\frac{1}{\sqrt{v}}({}_\infty\p^{2\a_n}+{}_\infty\bp^{2\a_n})\left(\sqrt{v}\,\,\cdot\,\right)\,,
\ee
where $m=2$ in Eqs.\ \Eq{lio} and \Eq{wey}:
\ba
({}_\infty\p^{2\a}+{}_\infty\bp^{2\a})f(x) &=& \frac{1}{\Gamma(2-2\a)}\int_{-\infty}^{+\infty}\rmd x'\, \left[\frac{\theta(x-x')}{(x-x')^{2\a-1}}+\frac{\theta(x'-x)}{(x'-x)^{2\a-1}}\right]\p_{x'}^2 f(x')\nonumber\\
&=&\frac{1}{\Gamma(2-2\a)}\int_{-\infty}^{+\infty}\frac{\rmd x'}{|x-x'|^{2\a-1}}\p_{x'}^2 f(x')\,.
\ea
At the classical level, the great advantage of \Eq{exms2} is that, in the single-scale case, it consists of just three terms $\p^2+(g_*/2)({}_\infty\p^{2\a}+{}_\infty\bp^{2\a})$ (again, weight factors are ignored) instead of seven. However, this $\cK$ is not quadratic, since $\bar\cD^{2\a}\neq\bar\cD^\a\bar\cD^\a$, which can lead to problems when quantizing the theory in Hamiltonian formalism: the kinetic term is not the square of a momentum operator. 

At present, it is not clear which definition between Eqs.\ \Eq{exms} and \Eq{exms2} will be more viable in the long run. They differ only in transient terms that can be dropped both at large and small scales, so that classically they give rise to the same physics. However, both have the added inconvenience of leading to a virtually symmetryless dynamics \cite{revmu}, a further point of concern if we want to do field theory and gravity with this formalism.


\subsection{Implicit multiscaling}

The multiscaling characterizing Eq.\ \Eq{exms} is of a twofold nature, an explicit one in the sum over $\a_n$ and an implicit one in the measure weight $v(x)$. These two structures have been combined independently and we imposed that the sum over $\a_n$ in the combination of fractional derivatives is the same sum over $\a_n$ inside $v(x)$. There is nothing wrong with this construction, but there may be a more elegant formulation where the scale hierarchy is all included within the measure $q(x)$ \cite{revmu}. Noting that the denominator $(x-x')^\a$ in Eqs.\ \Eq{lio} and \Eq{wey} for $m=1$ ($0<\a<1$) is the ultraviolet part of the profile in \Eq{S1d}, we can generalize those definitions as a left and right multifractional $q$-derivative:
\ba
{}_q\cD &:=& \int_{-\infty}^{+\infty}\rmd x'\,\frac{\theta(x-x')}{q(x-x')}\frac{\p}{\p x'}\,,\label{lio2}\\
{}_q\bar\cD &:=& \int_{-\infty}^{+\infty}\rmd x'\,\frac{\theta(x'-x)}{q(x-x')}\frac{\p}{\p x'}\,,\label{wey2}
\ea
where, again, the profile $q(x)$ is uniquely given by Eq.\ \Eq{qom}. These expressions are similar to the so-called variable-order fractional derivatives proposed by Lorenzo and Hartley \cite{LoH02}, although in our case $q(x-x')$ is completely fixed.

The kinetic operator in Eq.\ \Eq{S1d} is then
\be
\cK=\frac{1}{\sqrt{v}} \left[\frac12({}_q\cD-{}_q\bar\cD)\right]^2\left(\sqrt{v}\,\,\cdot\,\right)\,.
\ee
To understand the dynamics, we first need to spell out the properties of these derivatives. At short scales, $q(x-x')\sim |x-x'|^\a$ and Eqs.\ \Eq{lio2} and \Eq{wey2} reduce to the Liouville and Weyl derivatives, respectively:
\be
\text{small scales ($\ell\ll\ell_*$):}\qquad {}_q\cD\sim{}_\infty\p^\a\,,\qquad {}_q\bar\cD\sim{}_\infty\bp^\a\,,
\ee
while at large scales $q(x-x')\sim x-x'$ and Eqs.\ \Eq{lio2} and \Eq{wey2} give
\be
\text{large scales ($\ell\gg\ell_*$):}\qquad{}_q\cD\simeq \p\,,\qquad {}_q\bar\cD\simeq -\p\,.
\ee
We have not made a formal proof of these statements, but it should not be difficult. The Leibniz and integration-by-parts rules are also unknown but they should coincide with those of Weyl and Liouville fractional derivatives in the limit of small scales or in any plateau region of dimensional flow ($q\sim x^{\a_n}$). 

Therefore, we expect a complicated Leibniz and integration-by-parts rules everywhere at all scales of dimensional flow, except in plateau regions where a clean integration by parts of the type \Eq{ibp} emerges. For this reason, a variational principle valid at all scales may be ill defined in this case and an exact form of the equations of motion may be out of reach, although their asymptotic form at plateaux is obviously given by the limit of \Eq{cd2} for one exponent $\a$.

These considerations could eventually select the multifractional derivatives with explicit multiscaling as a simpler tool in quantum gravity, since they yield exact equations of motion.


\subsection{Multiscale differentials}

A third alternative is to introduce a multiscale differential based on the geometric coordinate \Eq{qom} or its simplified version in \Eq{S1d} \cite{revmu}:
\be\label{difq}
\bd q(x)=q(\rmd x)\,,
\ee
which is a linear combination of the usual and fractional \cite{frc1} differentials, $\bd q\sim\bd x+\bd |x|^\a+\ldots=\rmd x+|\rmd x|^\a+\dots$. (In $D$ dimensions, this differential generates the line element $\bd q(s)=\sqrt{g_{\mu\nu} \bd q^\mu(x^\mu)\otimes\bd q^\nu(x^\nu)}=q(\rmd s)= \sqrt{g_{\mu\nu} q^\mu(\rmd x^\mu)\otimes q^\nu(\rmd x^\nu)}$, where $g_{\mu\nu}$ is the metric.) The operator $\mathbbm{D}$ is a superposition of ordinary and fractional derivatives of the form (to be taken as indicative; coefficients are ignored)
\be
\bd=\bd q\mathbbm{D}\sim\rmd x\,\p+|\rmd x|^\a\p^\a+\ldots\,.
\ee
The following multiscale derivative and Laplacian are then defined implicitly:
\be\label{capuq2}
\mathbbm{D}:=\frac{\bd}{\bd q}\,,\qquad \cK=\mathbbm{D}^2\,.
\ee
These operators are invariant under translations, since $\mathbbm{D}_{x-\bar x}=\mathbbm{D}_x$, while $\cK$ is invariant also by ``$q$-boosts'' \cite{revmu}. Therefore, this theory has more symmetries than the theories with multifractional derivatives with explicit or implicit multiscaling.

In any plateau of dimensional flow, $\bd q\simeq(\rmd x)^{\a_n}$ and $\mathbbm{D}\simeq \frac{\bd}{(\rmd x)^{\a_n}}=\tilde\cD^{\a_n}$. Notice that $\mathbbm{D}\simeq \p_q$ in the near-infrared limit $\bd\to\rmd$ where the nonlinear part of $q$ is subdominant. Therefore, the theory with $q$-derivatives can be regarded as an approximation of the theory with multiscale derivatives (and, presumably, also of the other two theories with fractional derivatives) when the anomalous scaling effects are weak. The experimental constraints on the scales and parameters of the theory with $q$-derivatives might thus miss some important effects present in the fractional versions of the multiscale paradigm.

To determine the Leibniz and integration-by-parts rules, one should first define the operator $\mathbbm{D}$ appearing in the differential $\bd=\bd q\mathbbm{D}$. Again, we expect these rules to reduce to the usual ones in the infrared and to those of fractional derivatives in the ultraviolet. Since the operator \Eq{exms} with explicit multiscaling is already a well-defined linear combination of fractional derivatives, we reach the same conclusion of the previous section, namely, that the operator \Eq{exms} may be the best candidate for the concrete realization of multifractional theories with fractional derivatives. However, the main problem of the definitions \Eq{difq} and \Eq{capuq2} is that they are too abstract, which is the reason why we used qualitative expressions marked by ``$\sim$.'' Understanding their actual properties will require more work.


\section{Conclusions}

In this paper, we have further analyzed the proposals of \cite{revmu} for a multifractional calculus with viable applications to field theory and gravity. Without the pretense of being rigorous, we have considered some properties of scale-dependent derivative operators which, in physical applications to quantum gravity, are interpreted to encode the multiscaling of the underlying anomalous geometry. The conclusion is that the most promising multifractional theory possibly is the one with explicit multiscaling in fractional derivatives. However, only a full systematic study of the properties of all these operators will be able to confirm which is the most viable Laplacian from a theoretical and practical point of view. The value of the complex coefficients in \Eq{tilD0} will be especially important to determine a well-defined calculus and spectral theory \cite{frc4}. We will analyze the associated dynamics in detail in a future publication.


\section*{Acknowledgments}

\noindent{The author is under a Ram\'on y Cajal contract and is supported by the I+D grants FIS2014-54800-C2-2-P and FIS2017-86497-C2-2-P.}


\end{document}